# TERRESTRIAL EFFECTS OF NEARBY SUPERNOVAE IN THE EARLY PLEISTOCENE

B. C. Thomas[1], E. E. Engler[1], M. Kachelrieß[2], A. L. Melott[3], A. C. Overholt[4], and D.V. Semikoz[5,6]

1. Department of Physics and Astronomy, Washburn University, Topeka, Kansas 66621 USA;  brian.thomas@washburn.edu

2. Institutt for fysikk, NTNU, Trondheim, Norway

3. Department of Physics and Astronomy, University of Kansas, Lawrence, Kansas 66045 USA

4. Department of Science and Mathematics, MidAmerica Nazarene University, Olathe, Kansas, 66062 USA

5. APC, Universite Paris Diderot, CNRS/IN2P3, CEA/IRFU, Observatoire de Paris, Sorbonne Paris Cite, 119 75205 Paris, France

6. National Research Nuclear University "MEPHI" (Moscow Engineering Physics Institute), Kashirskoe highway 31, Moscow, 115409, Russia

ABSTRACT

Recent results have strongly confirmed that multiple supernovae happened at distances ~100 pc consisting of two main events: one at 1.7 to 3.2 million years ago, and the other at 6.5 to 8.7 million years ago. These events are said to be responsible for excavating the Local Bubble in the interstellar medium and depositing $^{60}$Fe on Earth and the Moon. Other events are indicated by effects in the local cosmic ray (CR) spectrum. Given this updated and refined picture, we ask whether such supernovae are expected to have had substantial effects on the terrestrial atmosphere and biota. In a first cut at the most probable cases, combining photon and cosmic ray effects, we find that a supernova at 100 pc can have only a small effect on terrestrial organisms from visible light and that chemical changes such as ozone depletion are weak.  However, tropospheric ionization right down to the ground due to the penetration of ≥TeV cosmic rays will increase by nearly an order of magnitude for thousands of years, and irradiation by muons on the ground and in the upper ocean will increase 20-fold, which will approximately triple the overall radiation load on terrestrial organisms. Such irradiation has been linked to possible changes in climate and increased cancer and mutation rates. This may be related to a minor mass extinction around the Pliocene-Pleistocene boundary, and further research on the effects is needed.

1. INTRODUCTION

There has long been speculation (Schindewolf 1954) that terrestrial mass extinctions (Bambach 2006; Melott & Bambach 2014) might be related to supernovae, but of necessity, such discussions of possible supernova (SN)-initiated events lacked precision. No mass extinction can be definitively linked to a SN, although circumstantial evidence in the form of agreement of geographical extinction patterns with expectations of atmospheric UV transmission has been presented for one possible link to a radiation event (Melott & Thomas 2009). Using available information, Gehrels et al. (2003) modeled effects of a hypothetical nearby SN, and found that a very close event (within ~8 pc) is necessary for a major mass extinction. Theirs was the only past computation to consider effects of both prompt photons and later CRs (modeled by a scaled-up SN 1987A).

We use the expanded knowledge of SN irradiation available from observations since then. Considerable improvement has been made in computational models of atmospheric ionization (Usoskin et al. 2011; Atri et al. 2010; Melott et al. 2016) and radiation effects on the ground from CRs (Overholt et al. 2015). We now have new parameters for actual recent SNe at a time when the terrestrial environment is reasonably well known. As we shall see, this event is not close enough to have precipitated a major mass extinction, but may have had noticeable effects. Given evidence for multiple events, these effects may have been additive.

A game-changing series of results began when Ellis et al. (1996) looked at live radioisotopes from moderately nearby SNe. A variety of radioisotopes were found to be excellent candidate markers for nearby SNe, as dust is transported to the solar neighborhood.

The first detection of terrestrial isotopes linked to a SN came from Knie et al. (1999) who found $^{60}$Fe in deep ocean ferromanganese crust using high resolution accelerator mass spectrometry. At that time there was also speculation that the formation of the hot, low-density Local Bubble in the interstellar medium might be connected with a chain of SNe. The second piece of evidence came from the idea that the CR spectrum we presently observe might show evidence of flux from one or more SNe (e.g. Erlykin and Wolfendale 2010). Recently, new experimental evidence and computational results have emerged which point to multiple SNe at a distance of about 100 pc beginning ~9 Ma (Ma = "Myr ago").

Our models are based on a "most likely" choice of event types based on the available data. Kachelrieß et al. (2015) found that an excess of positrons and antiprotons above about 20 GeV and the discrepancy in the slopes of the spectra of CR protons and heavier nuclei in the TeV-PeV energy range can be explained by a SN within a few hundreds of pc about 2 Ma. Such a source also may explain the observed plateau in the CR dipole anisotropy (Savchenko et al. 2015). Fry et al. (2015) conducted the most current assessment of $^{60}$Fe transport and deposition given information available then, and reached conclusions suggesting that an electron-capture SN (ECSN) at 100 pc about 2.2 Ma is the best fit to the experimental results to that date. For definiteness, we choose IIP as the most common ECSN type with the added advantage of fairly uniform properties. We model a single event, noting that results may be additive—except that the prior formation of the Local Bubble would be essential to the high flux of CR evident in one of our case studies.

More recently, results have appeared which further elaborate the picture. Breitschwerdt et al. (2016) looked at the $^{60}$Fe deposition history and trajectories of stars in the Local Bubble. They found that the nearest explosions were at ~100 pc, ~2 Ma, showing a consistent picture of the evolution of the Local Bubble and the $^{60}$Fe deposition in the crust. Wallner et al. (2016) presented a greatly expanded set of data from three different deep-sea archives, including four sediment cores, two iron-manganese crusts and two iron-manganese nodules. This dramatically expanded the amount of available $^{60}$Fe data and confirmed the same broad picture: two SN events within 100 pc, or possibly isotope transport from multiple SNe via the moving interstellar medium. The same parameters are indicated by the closely following publication of the detection of $^{60}$Fe in lunar samples (Fimiani et al. 2016) and the general chronology supported by recent direct detection in space (Binns et al. 2016).

At this point, it is useful to take a look at the effects on the Earth of both photons and CRs with the indicated sets of parameters. Results from CRs may vary dramatically depending upon the assumed magnetic field between the source and the Earth, but the contribution of local sources to the total CR intensity observed at present can be fixed (e.g. Savchenko et al. 2015).

There is a minor mass extinction about 2.59 Ma (Bambach 2006, Melott & Bambach 2014; note the revised geologic timescale between these publications.) This faunal change was not strongly pulsed (e.g. Behrensmeyer et al. 1997). Part of it is purely regional and associated with the closing of Panama, but there is some extinction connected with the generally cooling climate, which resulted in a transition to the Pleistocene period, with subsequent frequent glaciations. The atmospheric ionization effects we find are strong. It has been speculated that some of these effects may related to the rate of lightning strokes (Erlykin and Wolfendale 2010) and other changes that may alter climate (Mironova et al. 2015). We will consider (a) radiation on the ground and in the upper ~1 km of ocean and (b) ionization of the atmosphere. We will provide ionization data, which can later be used for questions of climate.

For now, we will treat three cases: cases A and B (see below) are based on an event at 300 pc, modeled as in Kachelrieß et al. (2015). Isotopic data suggest a closer event at 100 pc, which we model as case C. In all cases, we are interested in the most recent events ~2.5 Ma. Atmospheric ionization work is described in section 3.2, and air shower modeling to derive muon and neutron fluxes at ground level is described in section 3.3.

## 2. EFFECTS OF PHOTONS

### 2.1 GAMMAS AND X-RAYS

This section is based on case C events only. We shall see that at 100 pc the effects of photons are at most barely significant, so at 300 pc (as in cases A and B) they would be negligible.

There are no detections of gamma ray photons from Type IIP SNe. A type IIP is thought to be an explosion inside an extended envelope, which converts much of the energy to visible light. We therefore assume no significant prompt gamma-ray photons. There are some gammas from the remnant (Xing 2016), but this flux is far too small to be significant, $\sim 10^{-7}$ W m$^{-2}$. The threshold for significant atmospheric ionization is $\sim 10^4$ J m$^{-2}$ (Thomas et al. 2005). Over the first 3 years, the energy fluence would be three orders of magnitude smaller than this threshold, and for times greater than 3 years, atmospheric recovery becomes significant (Ejzak et al. 2007).

There are positive detections of X-rays in type IIP, but they indicate that IIP are uniformly X-ray underluminous (Dwarkadas 2014; Chakraborti et al. 2016). Again the flux at 100 pc would be $\sim 10^{-7}$ W m$^{-2}$. We conclude that we do not need to model photons of these energies.

### 2.2 VISIBLE and UV

There is a large body of evidence indicating that enhanced illumination at night is detrimental in a number of ways to many different types of organisms. Exposure to light during a normally dark period disrupts circadian rhythms, leading to loss of sleep, fatigue, changes in behavior, and other impacts detrimental to individuals and populations (Foster & Kreitzmann 2004; Dominoni et al. 2016 and references therein). Nighttime exposure to artificial lighting, particularly wavelengths between 400 nm and 500 nm, has been linked to cancer, suppression of immune systems, and reduced melatonin production. Brainard et al. (1984) found that irradiance of 0.186 µW cm$^{-2}$ of "cool white" fluorescent light caused a decrease in pineal melatonin production in the

Syrian hamster (*Mesocricetus auratus*), and that natural full moonlight irradiance as low as 0.045 µW cm$^{-2}$ caused pineal melatonin suppression in some animals, but not all. Changes in tree frog foraging behavior have been observed at illuminations above 10$^{-3}$ µW cm$^{-2}$ (Longcore & Rich 2004). Other research has shown that substantially larger fluxes can induce melatonin suppression in humans.

We examined available spectra of several Type IIP SNe, including 1999em (Hamuy et al. 2001), 2005cs (Bufano et al. 2009), SN2012aw (Bayless et al. 2013), 2013ej (Huang et al. 2015); spectrum data were downloaded from the Weizmann Interactive SN data REPository (http://wiserep.weizmann.ac.il/). After scaling each spectrum to represent a SN at 100 pc (case C), we find that the irradiance between 400 and 500 nm is roughly 0.03 to 0.06 µW cm$^{-2}$. This is comparable to the irradiance values shown to have some effect in animals, but significantly smaller than the values used in human melatonin suppression studies.

Spectra available for SN2013ej and SN2012aw extend to the hard UV. Scaled to 100 pc distance, the irradiance would be approximately 10$^{-3}$ W m$^{-2}$ at the top of the atmosphere. This is several orders of magnitude smaller than the Solar UV irradiance at the top of the atmosphere and so is unlikely to have any significant impact on the atmosphere or significantly increase UV irradiance at Earth's surface.

To summarize: In case C, some physiological effects of blue light in the night sky might be significant for a few weeks. Otherwise visible and UV photons are negligible. This is in agreement with the estimates of Ellis and Schramm (1995), who did not consider the effects of blue light.

## 3. EFFECTS OF CRs

### 3.1 SUMMARY OF PROPAGATION

The propagation of Galactic CRs is usually approximated as diffusion. In the limit that the turbulent magnetic field dominates, diffusion becomes isotropic and the diffusion tensor simplifies to $D(E)=D_0 (E/E_0)^{\alpha}$. In an alternative approach, one numerically calculates the trajectories of individual CRs solving the Lorentz equation in the turbulent and regular Galactic magnetic field (GMF). This approach is ideally suited to the study of nearby sources in the presence of a regular field: The CR density emitted by such sources is strongly anisotropic, because CRs propagate mainly along the regular field.

In the trajectory approach, we use the code described and tested in Giacinti et al. (2012) together with the GMF model of Jansson-Farrar (Jansson & Farrar 2012a, 2012b). The turbulent part of the field is chosen as isotropic Kolmogorov turbulence with $L_{max}$=25 pc as the maximal length of the fluctuations. For the numerical calculations, we use nested grids, allowing us to choose an effective $L_{min}$ sufficiently small compared to the Larmor radius of the CRs considered. The power-spectrum $P(k) \sim k^{-5/3}$ of the turbulent component was normalized to reproduce the observed B/C ratio, as discussed in Giancinti et al. (2015). The CR fluxes obtained in this model reproduce over a wide range of energies all available experimental data for individual groups of CR nuclei (Giancinti et al. 2015).

We are interested in the case of a relatively young source, T< few Myr. Since the propagation of CRs with energies below 100 TeV on Myr time scales is strongly anisotropic in the presence of a regular field, a sharp enhancement of the CR flux occurs if the source and observer are connected by a magnetic field line. In this case, the contribution of a single source can dominate the observed total CR intensity at the Earth as shown in Kachelrieß et al. (2015). Following this work, for cases A and B here, we place a source at the distance R = 300 pc and 50 pc from the GMF line passing through the Solar system. Note that our results are not very sensitive to the exact value of R as long as $R^2/T < D(\Theta)$, where $D(\Theta)$ is the diffusion coefficient towards the source. We assume that the source injected instantaneously CRs (E> 1 GeV) with spectrum $dN/dE \sim E^{-2.2} \exp(-E/E_c)$, cutoff $E_c$= 1 PeV, and total injection energy $E_{tot}$ = 2.5 x $10^{50}$ erg. We record the path length spent in a 50 pc sphere around the Earth, which can be converted to the local CR flux at a given time interval. The observed plateau of the CR dipole amplitude is for these parameters naturally explained (Savchenko et al. 2015), its decrease below TeV energies may be caused by the offset of Earth with respect to the magnetic field line going through the source. We account for the uncertainties in the propagation model, as e.g. the unknown off-set towards the connecting field line, by considering two basic scenarios: In case A, we assume that the resulting low-energy cutoff was absent at early times, while we use in case B a low-energy cutoff corresponding to a constant distance of the Solar System from the connecting field line.

An additional case is motivated by the recent $^{60}$Fe data, which suggest a nearer event, probably within the Local Bubble. We assume that it is the second event in an already-excavated region, as would be consistent with the detections of Wallner et al. (2016). We therefore assume a distance of order 100 pc and an intervening field of order 0.1 µGauss, dominated by its turbulent component (e.g. Avillez and Breitschwerdt 2005). Therefore we can use the diffusion approximation in case C with $D_0$ = 2 x $10^{28}$ cm$^2$s$^{-1}$ for $E_0$ = 1 GeV. In Figure 1, we show the high-energy CR spectra as a function of time incident on the Earth in each of these cases. Times are measured from the escape of CRs from the expanding SN remnant. Case B has negligible CR flux at all times.

In the limit

$$r \ll r_{diff} = 0.5 \text{pc} \left[ \left(\frac{t}{\text{yr}}\right) \left(\frac{E}{\text{GeV}}\right)^{1/3} \right]^{1/2} \tag{1}$$

the intensity is simply

$$E^2 \cdot I(E,r) = 4.4 \times 10^{14} \left(\frac{\text{eV}}{\text{cm}^2 \cdot \text{s} \cdot \text{sr}}\right) \left(\frac{E}{\text{GeV}}\right)^{-0.2} e^{-E/E_c} \left(\frac{\text{pc}}{r_{diff}}\right)^3 . \tag{2}$$

The formula for $I(E,r)$ from a bursting source in the general case may be found in Kachelrieß et al. (2015).

In Figure 1 we plot flux*$E^2$ so that the area under the curve is proportional to the total energy between limits. It can be seen that the major difference in some cases is the large amount of energy deposited by CRs in excess of a TeV. This is very different from solar events, and implies penetration and ionization at much greater atmospheric depths, near to the ground.

3.2 ATMOSPHERIC EFFECTS

Atmospheric ionization by CRs is treated as in Melott et al. (2016), by convolving the CR spectra (for SN cases and GCRs) shown in Fig. 1 with tables that give ionization as a function of altitude from Atri et al. (2010), for primaries between 10 and $10^6$ GeV. Our procedure does not take into account solar or geomagnetic rigidity, which should have only a small effect for primaries ≥ 10 GeV. Our results may be slightly too large above the lower stratosphere since lower energy primaries are more strongly affected by solar and geomagnetic rigidity and have a greater ionization effect at higher altitudes. However, comparison of our results with other published results (Jackman et al. 2016) shows good agreement within the variation caused by present day rigidity effects.

In Case A, the only substantial effect we have found is that ionization in the lower troposphere exceeds that from normal galactic CRs by about 70% (Figure 2a). This level could be expected to persist for about 20 kyr. Such ionization might cause an increase in lightning (Erlykin and Wofendate 2010). There is a wide variety of possible effects, including chemical changes, chemistry-dynamics feedbacks, the global electric circuit and cloud formation (Mironova et al. 2015). A major difference between our results and the usual ionization sources is that the high energy SN CRs ionize much more strongly down to the lower troposphere. Since stronger tropospheric ionization is unusual, its effects have not been well-studied or simulated.

Case B, with the blocking magnetic field, has no significant flux (see Figure 2b).

Case C, motivated by the recent $^{60}$Fe detections, gets a significant prompt flux at 500 yr, since the distance is smaller (than cases A and B), with a much weaker intervening magnetic field. The results, shown in Figure 2c, document a nearly order of magnitude increase in ionization, right down to the ground, 500 yr after escape of the CRs from the SN remnant. Effects such as chemical changes, chemistry-dynamics feedbacks, changes in the global electric circuit and cloud formation are beyond the scope of this paper, but there is now a strong motivation with definite numbers for investigating them. This strong ionization will taper off by 5 kyr after the SN. We cannot offer a conclusion, except some on chemical changes. Our results are summarized in Figure 2, and hope that they offer an opportunity for further research to clarify the effects.

Atmospheric ionization can have important impacts on chemistry. Odd-nitrogen oxides ($NO_y$) are produced, leading to destruction of stratospheric ozone and subsequent rain-out of $HNO_3$ (Thomas et al. 2005). While a full analysis is beyond the scope of this work, we will discuss some results from 2D (altitude-latitude) atmospheric chemistry modeling, completed using the Goddard Space Flight Center two-dimensional atmospheric chemistry and dynamics model (Thomas et al. 2005 and references therein). We find that for case C at 500 yr, in the long-term steady-state (achieved after about 10 years), $NO_y$ abundance in the upper troposphere is increased about 200%, compared to a typical pre-industrial background. We find globally averaged $O_3$ column density depletion of about 7% with locally higher depletion up to about 12% (primarily in the polar regions); consequently changes in UV transmission to the ground will be insignificant (Thomas et al. 2015). $NO_y$ produced by ionization is removed by rain/snow-out of $HNO_3$. At steady-state in our modeling the global annual average deposition above background is about $8 \times 10^{-3}$ g m$^{-2}$; this value is about 1.3 times that generated by lightning and other nonbiogenic sources in today's atmosphere (Schlesinger 1997). This does not appear to be enough to have major effects on the biota, because the changes in UV and nitrate deposition on the ground are not large.

3.3 RADIATION ON THE GROUND

The two species of particles that produce the largest amounts of ground level cosmogenic radiation are muons and neutrons. Although other species of cosmic ray secondaries are produced in the atmosphere, they do not reach ground level in numbers sufficient for measurable radiation dose. We based our work on tables constructed by extensive simulations using CORSIKA (COsmic Ray SImulations for KAscade) (Pierog et al. 2007) which is a Monte Carlo code used to study air showers generated by primaries up to 100 EeV. Neutron and muon fluxes at ground level were found by convolving the primary spectra with tables of Overholt et al. (2013) and Atri & Melott (2011), based on these shower simulations. Overholt et al. (2013) used MCPNx (Hagmann et al. 2007) to propagate the neutrons from the 50 MeV output of CORSIKA down through the atmosphere and down to thermal energies. In what follows, we use the procedures used in Overholt et al. (2013): Our Table I is based on neutron radiation dose "estimated by taking order of magnitude averages of ionizing radiation dose per neutron from Alberts et al. (2001) and multiplying by the number of neutrons found" in our procedure. "Muon radiation dose was found by calculating the energy loss of high energy muons traveling through matter. Assuming a radiation-weighting factor of 1, this provides an equivalent dose. Research suggests that this assumption is adequate for deeper organs, but greatly underestimates dose at skin level (Siiskonen 2008). For these calculations, we use the energy loss of muons traveling through water as an approximation for biologic matter (Klimushin et al. 2001)."

In Table 1, we show the muon and neutron doses on the ground for cases A, B, and C, as well as the present day background for cosmic ray primaries of energy >10 GeV. We must compare the increase due to muons and/or neutrons to the total background radiation dose. The radiation dose for cases A and B is significantly lower than the background radiation dose in most time periods, and as such is incapable of producing damage to terrestrial biota.

The most significant dose is achieved in early times of case C. In fact, the 20-fold enhancement of radiation from muons is enough to roughly triple the total radiation dose. This radiation dose is nearly double the average worldwide radiation background dose from all sources (2.4 mSv yr$^{-1}$, UNSC 2008), and would be likely to cause increased cancer risk and mutations. This is due in part to neutrons and muons having greater penetration than the majority of background radiation, which comes from inhalation of radon (UNSC 2008). One can qualitatively characterize the dose level "as if every organism got a CT scan, every year." This is not disastrous, but might be noticeable in the fossil record. While most of the normal background is roughly proportional to surface area of the organism, muon irradiation is, due to penetration, proportional to volume. Thus, the relative change in irradiation will be greatest for large organisms. Since muons can penetrate up to about a kilometer in water, all but deep ocean organisms would be affected.

3.4 DIRECT DEPOSITION AND COSMOGENIC ISOTOPES

We must consider the possibility of radioactivity on the ground. As discussed by Ellis et al. (1996), there are two primary sources.

Cosmogenic isotopes such as $^{10}$Be and $^{14}$C are generated by the actions of CRs in the atmosphere. Using the discussion of Ellis et al. (1996) and considering a source at ~100 pc, we would expect ~$10^5$ atoms cm$^{-2}$ of radionuclides during the passage of CRs generated by the event. This is consistent with our own estimates of the enhancement of CRs in case C. Cases A and B would be negligible in this regard.

Direct deposit of SN-generated isotopes from a ~10 M$_\odot$ event at 100 pc can be expected, according to Ellis et al. (1996) to deposit at most of order $10^6$ atoms cm$^{-2}$ assuming that the interstellar medium has been previously cleared by another event; if not they estimate that the blast wave would stall before reaching 100 pc. These amounts are too small to be significant for the environment, but the direct deposit amounts lie at the base of detection (e.g. Wallner et al. 2016). Cases A and B are not based on isotope deposition, and should be too far away for a significant effect.

## 4. SUMMARY OF EFFECTS AND DISCUSSION

In case A, we found a modest (~70%) increase in tropospheric ionization which would persist for ~20 kyr. In case C, we found small effects due to enhanced blue light (which would persist for a few weeks) and a major—order of magnitude—increase in tropospheric ionization which would persist for at least 1000 yr. It is possible that this could trigger climate change, especially if instability was already present. Past work that considered the effect of cosmic ray fluctuations on climate (e.g. Mironova et al. 2015) did not consider an order of magnitude increase in ionization down to the bottom of the troposphere, made possible here by the presence of large numbers of high energy CRs (Figure 1c). This deserves more attention. Muon irradiation on the ground will increase by a factor of about 20 in case C, tripling the overall radiation dose. These effects are different than the ozone depletion/UV increase commonly associated with astrophysical radiation events. While not catastrophic, the effects may be detectable in the fossil record and could possibly be related to the secondary mass extinction noted at the Pliocene-Pleistocene boundary.

BCT, ALM, and ACO gratefully acknowledge the support of NASA Exobiology grant NNX14AK22G. We have had useful comments from J. Beacom, D. Breitschwerdt, B. Fields, R. Mandel, M. Medvedev, A. West and two referees. Computation time was provided by the High Performance Computing Environment (HiPACE) at Washburn University; thanks to Steve Black for assistance.

REFERENCES

Alberts, W. G. et al. 2001, J. of the ICRU, 1(3).

Atri, D. & Melott, A.L. 2011, Rad. Phys. Chem. 81, 701.

Atri. D. et al. 2010, JCAP, 008, doi:10.1088/1475-7516/2010/05/008.

Avillez, M.A. & Breitschwerdt, D. 2005 A&A 436, 585. DOI: 10.1051/0004-6361:20042146

Bambach, R.K. 2006, Annu. Rev. Earth Planet. Sci. 34, 127.Bayless, A.J. et al. 2013, ApJL, 764, L13.

Behrensmeyer, A.K. et al. 1997, Science 278, 1590.

Binns, W. R. et al. 2016, Science 352, 677-680. DOI:10.1126/science.aad6004

Brainard, G.C. et al. 1984, J. Pineal Res., 1, 105.

Breitschwerdt, D. et al. 2016, Nature 532, 73-76.

Bufano et al. 2009, ApJ, 700, 1456.

Chakraborti, S. et al., 2016, ApJ 817, 22. http://dx.doi.org/10.3847/0004-637X/817/1/22

Dominoni, D.M. et al. 2016, Biol. Lett., 12, 20160015.

Dwarkadas, V.V. 2014, MNRAS 440, 1917.

Ejzak, L.M. et al., 2007, ApJ 654, 373.

Ellis, J., & Schramm, D.N., 1995 PNAS 92, 235.

Ellis, J. et al. 1996, ApJ 470, 1227.

Erlykin, D. & Wolfendale, A.W. 2010, Surv. Geophys. 31, 383.

Fimiani, L. et al. 2016, Phys. Rev. Lett. 116. DOI: 10.1103/PhysRevLett.116.151104

Foster, R.G. & Kreitzmann, L. 2004, Rhythms of life: the biological clocks that control the daily lives of every living thing (New Haven, CT: Yale U. Press)

Fry, B. J. et al. 2015, ApJ 800, 71.

Gehrels, N. et al. 2003, ApJ 585, 1169.

Giacinti, G., Kachelrieß, M., Semikoz, D.V., Sigl, G. 2012, JCAP 1207, 031.

Giacinti, G., Kachelrieß, M., Semikoz, D.V. 2015, Phys. Rev. D 91, 083009.

Hagmann, C. et al. 2007, Monte Carlo simulation of proton-induced cosmic-ray cascades in the atmosphere, Tech. Rep., UCRL-TM-229452, Lawrence Livermore Natl. Lab., Livermore, Calif.

Hamuy, M. et al. 2001, ApJ, 558, 615.

Huang, F. et al. 2015, ApJ, 807, 59.

Jackman, C.H. et al. 2016, Atmos. Chem. Phys. 16, 5853.

Jansson, R. & Farrar, G.R. 2012a, ApJ 757, 14.

Jansson, R. & Farrar, G.R. 2012b, ApJ 761, L11.

Kachelrieß, M. et al. 2015, Phys. Rev. Lett. 115, 181103.

Klimushin, S., Bugaev, E., Sokalski, I., 2001, Proc. of the 27$^{th}$ Intl. CR Conf. 07-15 August, 2001. Hamburg, Germany, 1009.

Knie, K. et al. 1999, Phys. Rev. Lett. 83, 18. Doi:10.1103/PhysRevLett.83.18

Longcore, T. & Rich, C. 2004, Front. Ecol. Environ. 2, 191-198.

Melott, A.L. & Bambach, R.K. 2013, Paleobiology 40, 177.

Melott, A.L. & Thomas, B.C. 2009, Paleobiology 35, 311. doi:10.1666/0094-8373-35.3.311

Melott, A.L. et al. 2016, JGR Atmospheres 121. DOI: 10.1002/2015JD024064.

Mironova, I.A. et al. 2015, Spa. Sci. Rev. 194, 1. DOI 10.1007/s11214-015-0185-4

Overholt, A.C. et al. 2013, JGR Space Physics 118, 3765. doi:10.1002/jgra.50377

Overholt, A.C. et al. 2015, JGR Space Physics 120. doi:10.1002/2014JA020681.

Pierog, T. et al. 2007, Prepared for 30th International Cosmic Ray Conference. Latest Results from the Air Shower Simulation Programs CORSIKA and CONEX. (ICRC 2007), Merida, Yucatan, Mexico arXiv:0802.1262v1.

Savchenko, V., Kachelrieß, M. & Semikoz, D.V. 2015, Astrophys. J. 809, L23.

Schindewolf, O. H. 1954. Neues Jb. Geol. Paläontol. 10, 457–465.Schlesinger, W.H. 1997, Biogeochemistry (2$^{nd}$ ed.; San Diego: Academic)

Siiskonen, T., 2008, Radiation Prot. Dos., 128(2), 234.

Thomas, B. C. et al. 2005, ApJ 634, 509.

Thomas, B.C. et al. 2015, Astrobiology 15, 207.

UNSC on the Effects of Atomic Radiation 2008. New York: United Nations, 4. ISBN 978-92-1-142274-0

Usoskin, I.G. et al. 2011, Atmos. Chem. Phys., 11, 1979-2011, doi:10.5194/acp-11-1979-2011.

Wallner, A. et al. 2016, Nature 532, 69-72.

Xing, Y. et al. 2016, arXiv:1603.00998 [astro-ph.HE]

|  | Case A | | Case B | | Case C | |
| --- | --- | --- | --- | --- | --- | --- |
|  | Muons (mSv) | Neutrons (mSv) | Muons (mSv) | Neutrons (mSv) | Muons (mSv) | Neutrons (mSv) |
| 2 Myr | 0.0229 | 1.13E-05 | 0.029 | 1.13E-05 | 2.99E-04 | 1.27E-06 |
| 500 kyr | 0.0832 | 2.50E-05 | 0.0205 | 1.61E-06 | 2.26E-03 | 6.62E-06 |
| 100 kyr | 0.204 | 2.37E-05 | 3.33E-03 | 1.06E-07 | 0.0221 | 5.99E-05 |
| 20 kyr | 0.553 | 4.68E-05 | 8.18E-06 | 2.50E-10 | 0.183 | 4.18E-04 |
| 5 kyr | 0.259 | 1.28E-05 | 1.81E-14 | 7.35E-19 | 0.876 | 1.44E-03 |
| 500 yr | -- | -- | -- | -- | 4.02 | 1.51E-03 |
|  |  | | Present Day | | | |
|  |  | | Muons (mSv) | Neutrons (mSv) | | |
|  |  | | 0.201 | 0.166 | | |

Table 1
Annual ground level secondary radiation dose, in units of mSv, due to muons and neutrons, for our three SN cases at six different times, measured from the release of cosmic rays from the SN remnant. Also shown for comparison is the present day annual dose due to muons and neutrons. Case A is based on a 300 pc supernova with no set-off from the magnetic field line, and case B the same, with a set-off equivalent to that of the present day. Case C is based on a 100 pc supernova as indicated by the $^{60}$Fe results, inside the Local Bubble so that only a weak turbulent magnetic field intervenes between the Solar System and the supernova. Case C, 500 yr, might induce significant effects on terrestrial and upper-ocean organisms.

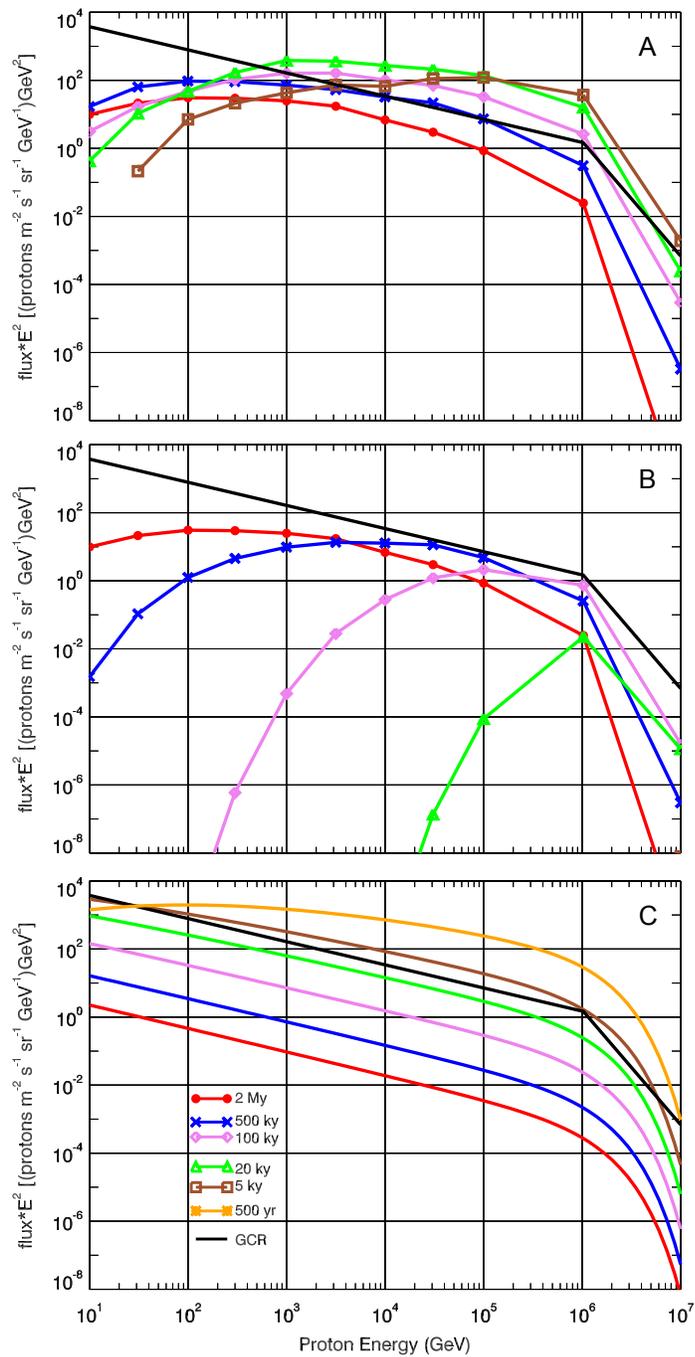

Figure 1

Cosmic ray flux spectrum (times $E^2$) for each case (A, B, C) at several times (red = 2 My, blue = 500 ky, purple = 100 ky, green = 20 ky, brown = 5 ky, orange = 500 yr), along with Galactic cosmic ray background flux (GCR; black line). Symbols are omitted from case C due to higher resolution in energy used in modeling.

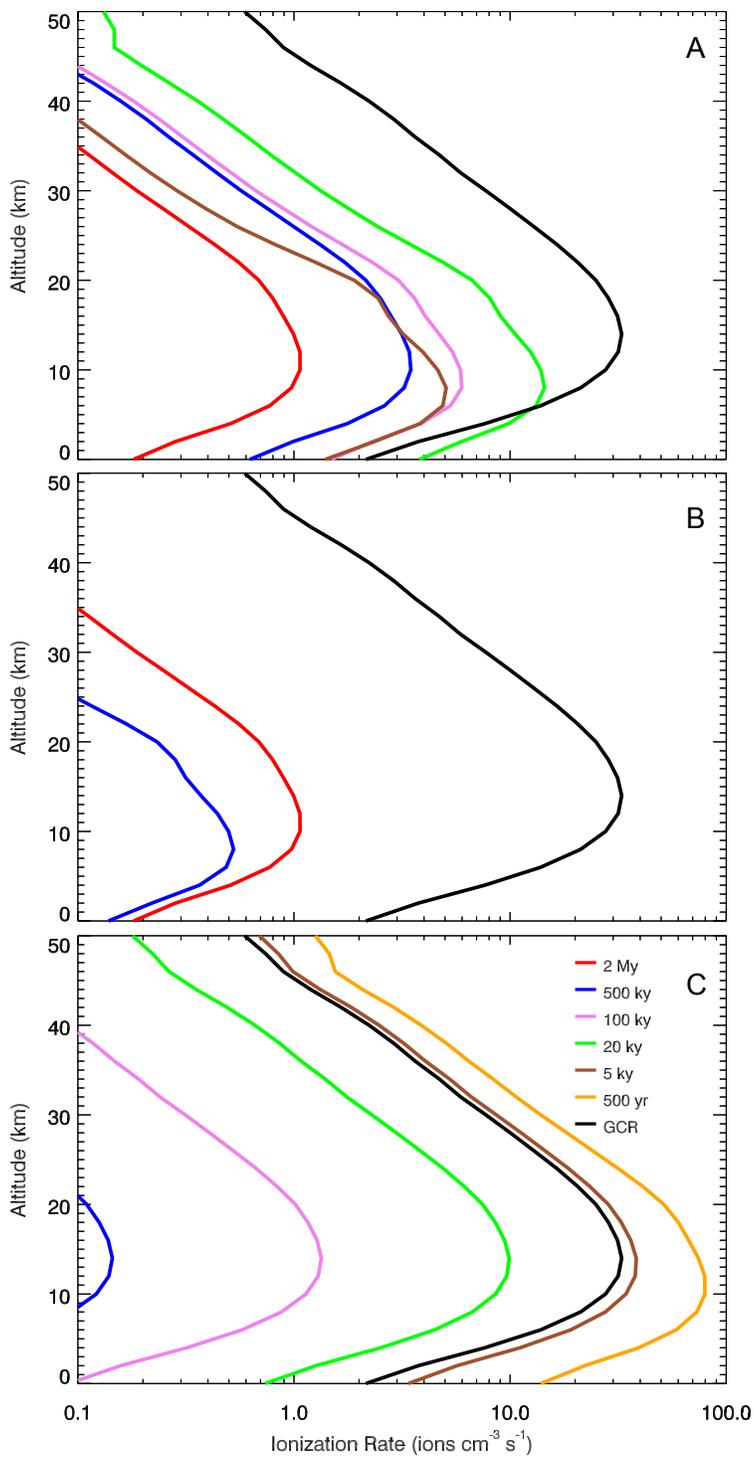

Figure 2

Atmospheric ionization rates for each case (A, B, C) at several times (red = 2 My, blue = 500 ky, purple = 100 ky, green = 20 ky, brown = 5 ky, orange = 500 yr), along with Galactic cosmic ray background ionization rate (GCR; black line).